\documentclass[
    ,final            
  ]
  {aipproc}

\layoutstyle{6x9}

\begin{document}

\title{Understanding the Forward Muon Deficit in Coherent Pion Production}

\classification{}
\keywords      {}

\author{L. M. Sehgal}{
  address={Institute of Theoretical Physics (E), RWTH Aachen, 52056 Aachen, Germany}
}



\begin{abstract}
For any inelastic process $v_{\ell} + I \to \ell^- + F$
with $m_{\ell} = 0$, the cross section at $\theta_{\ell} = 0$ is given by Adler's PCAC theorem.
Inclusion of the lepton mass has a dynamical effect (``PCAC-screening'')
caused by interference of spin-zero ($\pi^+$) and spin-one exchanges.
This effect may be relevant to the forward suppression reported in recent experiments.\\
\end{abstract}

\maketitle


\section{}
Recent experiments with low energy neutrino beams suggest
that in inelastic CC events, there are fewer muons coming out at small angles than expected.
 (For a review of the data, see the talk of Bonnie Fleming) \cite{B. Fleming}.
  he evidence comes principally from two-track events that are interpretable as
    $ \nu_{\mu} + (p,n) \to \mu^- + (p,n) + \pi^+$ with nucleon undetected, or coherent
     $\pi^+$ production $\nu_{\mu} +$ Nucleus $\to \mu^- + \pi^+ +$ Nucleus. In particular,
      the K2K experiment ($E_{\nu} \approx \, 1.3 GeV$) has reported a deficit at
       low $Q^2 (Q^2 < 0.1 GeV^2)$ which they interpret as a suppression of coherent
        $\pi^+$, obtaining an upper limit
         $\sigma(coh \, \pi^+) / \sigma(CC) < 0.6 \%$  \cite{M. Hasegawa et al},
          compared with a theoretically expected value of $2 \%$ \cite{ReinSehgal}.

The deficit is puzzling since there appears to be evidence for NC coherent
 $\pi^0$ production $\nu_{\mu} +$ Nucleus $\to \nu_{\mu} + \pi^0 +$
 Nucleus \cite{S. Nakayama} at roughly the expected rate
 $\sigma(coh \, \pi^0)/ \sigma(CC) \approx 1$ \%, and a ratio
  $ \sigma(coh \, \pi^+)/\sigma(coh \, \pi^0) =2 $ is expected from fairly general
   isospin considerations.

This situation has prompted us to ask whether the deficit could be a dynamical
 effect caused by the nonzero mass of the muon in the CC channel, absent in the NC process.
  We recall that in any inelastic CC reaction $\nu_{\mu}+ I \to \mu^- + F, F \neq I$,
   the cross section in the forward scattering configuration for $m_{\ell} = 0 $ is
    predicted by Adler's PCAC theorem \cite{S. L. Adler}
\begin{equation}
\left( \frac{d \sigma}{d x dy}\right)_{\theta =0} = \frac{G^2 ME_{\nu}}{\pi^2} f^2_{\pi} (1-y)
 \sigma (\pi^+ + I \to F) |_{E_{\pi}= E_{\nu}y}
\label{md1}
\end{equation}
For non-forward scattering, this result is expected to be modified by a slowly-varying
factor $(1 +Q^2/M^2_A)^{-2}$, where $M_A \approx 1 GeV$ is the typical mass of the spin-one
 $(1^{++})$ mesons mediating the process. If the muon mass is not neglected, however,
  the process receives an additional contribution due to the exchange of a spin-zero
   $ \pi^+$ meson. It was shown by Adler  \cite{S. L. Adler} that the forward theorem
    is modified by a multiplicative factor, which may be written as \cite{ReinSehgal06}
\begin{equation}
C_{Adler} = \left( 1 - \frac{1}{2} \frac{Q^2_{min}}{Q^2 + m^2_{\pi}}\right)^2
 + \frac{1}{4} y \frac{Q^2_{min}(Q^2-Q^2_{min})}{Q^2 +m^2_{\pi})^2}
\label{md2}
\end{equation}
where
\begin{equation}
Q^2_{min} = m^2_{\ell} \frac{y}{1-y}
\label{md3}
\end{equation}

This correction is valid for small angles, and contains the important terms in which the
factor $ m^2_{\mu}$ is accompanied by the pion propagator $1/(Q^2 + m^2_{\pi})$.
The factor $C_{Adler}$ has non trivial consequences for all inelastic cross
sections at small angles. For forward scattering, in particular,
\begin{equation}
C_{Adler} (\theta = 0) = \left( 1 - \frac{1}{2} \frac{Q^2_{min}}{Q^2_{min} + m^2_{\pi}}\right)^2
\label{md4}
\end{equation}
The minus sign within parentheses indicates that the effect of pion-exchange
 is a destructive interference. Taking an average value $y \approx 1/2$,
  the forward suppression factor is
\begin{equation}
C_{Adler} (\theta = 0, y = 1/2) = \left( 1 - \frac{1}{2} \frac{ m^2_{\mu}}{ m^2_{\mu} + m^2_{\pi}}
\right)^2 = 70 \% !
\end{equation}

\indent We have investigated  \cite{ReinSehgal06},
 the consequences of this screening effect in the coherent process
  $ \nu_{\mu} + C^{12} \to \mu^- + \pi^+  +C^{12}$ using the model described
   in \cite{ReinSehgal}. The effects on $d \sigma /d \, cos \theta_{\mu}$
   and $d \sigma /d Q^2$  are shown in Fig. 1 and Fig. 2, and exhibit a forward muon deficit.
Note that a comparison of the $m_{\mu} \neq 0$ and $m_{\mu}=0$ cases is essentially
 a comparison of
$\nu_{\mu}$ and $ \nu_e$ scattering, and that a ``muon deficit''
could equally be regarded as an ``electron excess''.

In applying the above suppression mechanism to the K2K data,
 our analysis \cite{ReinSehgal06} indicates that the coherent
  $\pi^+$ signal in the domain $Q^2 < 0.1 GeV^2$ is suppressed by a factor
   $\langle C_{coh}\rangle \approx 0.77$. We have also estimated the incoherent
    resonant background, using the resonance model \cite{ReinSehgal81},
     and obtain a suppression factor $\langle C_{res}\rangle \approx 0.85$.
      These results allow a reinterpretation of the K2K deficit in the interval
       $Q^2 < 0.1 GeV^2$, and reduce the discrepancy to about $ 2 \sigma$.
        A detailed discussion of muon mass effects will appear in \cite{BergerSehgal}.

\begin{theacknowledgments}
I thank Chris Berger for help in preparing the figures.
Thanks to Jorge Morfin and the organizers of NuInt07 for
the invitation to present these ideas at their Workshop.
\end{theacknowledgments}

\vfill

\begin{figure}[h]
  \includegraphics[height=.35\textheight]{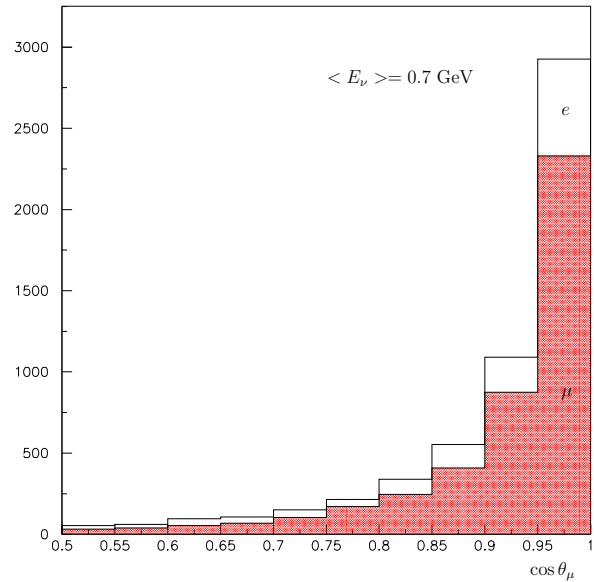}
  \caption{Muon mass corection in $d \sigma^{coh} /d \, cos \, \theta_{\mu}$ for MiniBoone energy$ \langle E_{\nu}\rangle = 0.7 \, GeV$}
\end{figure}

\begin{figure}[b]
  \includegraphics[height=.35\textheight]{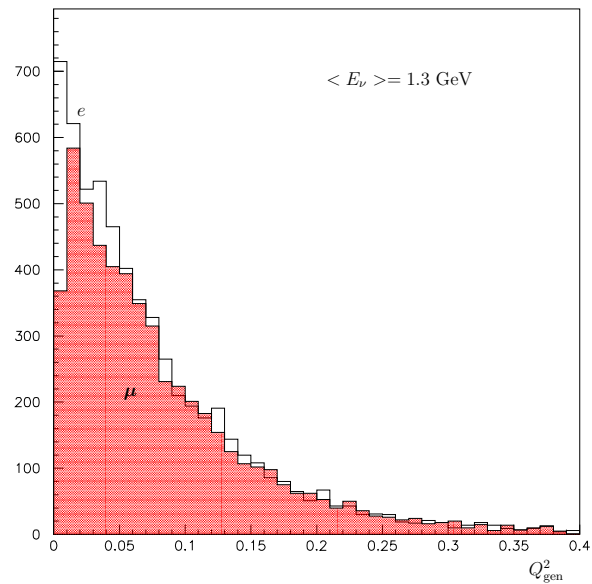}
  \caption{Muon mass correction in $d \sigma^{coh} / d Q^2$ for K2K energy $\langle E_{\nu}\rangle = 1.3 \, GeV$}
\end{figure}

\end{document}